\documentclass{article} 
\usepackage{iclr2020_conference,times}

\usepackage{hyperref}       
\usepackage{url}            
\usepackage{booktabs}       
\usepackage[pdftex]{graphicx}

\title{Smart Home Appliances:\\Chat with Your Fridge}

\author{Denis~Gudovskiy, Gyuri~Han\\
Panasonic $\beta$ AI Lab, Mountain View, CA, 94043 \\
\texttt{\{denis.gudovskiy, gyuri.han\}@us.panasonic.com} \\
\And
Takuya~Yamaguchi, Sotaro~Tsukizawa \\
Panasonic AI Solutions Center, Osaka, Japan, 571-0050 \\
\texttt{\{yamaguchi.takuya2015, tsukizawa.sotaro\}@jp.panasonic.com}
}

\iclrfinalcopy 
\begin{document}

\maketitle

\begin{abstract}
Current home appliances are capable to execute a limited number of voice commands such as turning devices on or off, adjusting music volume or light conditions. Recent progress in machine reasoning gives an opportunity to develop new types of conversational user interfaces for home appliances. In this paper, we apply state-of-the-art visual reasoning model and demonstrate that it is feasible to ask a \textit{smart fridge} about its contents and various properties of the food with close-to-natural conversation experience. Our visual reasoning model answers user questions about existence, count, category and freshness of each product by analyzing photos made by the image sensor inside the smart fridge. Users may chat with their fridge using off-the-shelf phone messenger while being away from home, for example, when shopping in the supermarket. We generate a visually realistic synthetic dataset to train machine learning reasoning model that achieves 95\% answer accuracy on test data. We present the results of initial user tests and discuss how we modify distribution of generated questions for model training based on human-in-the-loop guidance. We open source code for the whole system including dataset generation, reasoning model and demonstration scripts\footnote{\href{https://github.com/gudovskiy/beta-fridge}{Our code is available at github.com/gudovskiy/beta-fridge}}.
\end{abstract}

\section{Introduction}\label{sec:intro}
Recent achievements in the field of machine learning (ML) and, in particular, deep neural networks (DNNs) already make feasible some elements of the highly-anticipated general artificial intelligence (AI). For instance, DNN-based models surpassed previously achieved baselines or even human-level accuracy in many tasks including image classification~\citep{he}, object detection~\citep{mod_tradeoffs}, and instance segmentation~\citep{mask} in computer vision. Natural language processing (NLP) tasks such as language understanding~\citep{bert}, speech-to-text~\citep{wavenet}, and visual question answering (VQA)~\citep{vqa2} are progressing rapidly as well. A distinct example is the last task in which language processing is combined with computer vision and users are able to ask questions about visual pictures. Therefore, VQA can be considered as the first step towards more abstract visual reasoning or, in general, machine reasoning field.

Previous \textit{feature engineering} model paradigm is currently being replaced by a \textit{universal DNN architecture}, which is trained on the collected data with minimum developer effort. In addition, the nature of ML models allows to use any multimodal sensor data by fusing input modalities such as camera photos, radio-frequency signals, thermostat readings, and even language VQA questions for home applications. In this paper, we develop a \textit{visual reasoning} ML-based system for smart home appliances with camera and language modalities.

Data availability becomes a crucial aspect for ML model development. Though unsupervised learning made significant progress in certain areas~\citep{infomax}, DNN models are still mostly rely on traditional supervised learning. The data aspect of supervised ML usually consists of two main steps: collection of annotated data and subsequent dataset generation. The latter is involved with the design of relevant training dataset that matches distribution of actual user needs. We address data availability issue by generating a synthetic dataset called \textit{FRIDGR} for smart fridge application. We employ human-in-the-loop technique to match distribution of training dataset with the collected during user tests data.

Unlike conventional classification or regression models, DNN-based machine reasoning targets to perform more complicated tasks with higher level of abstraction and better knowledge generalization. VQA models~\citep{Yu_2019_CVPR} are typically capable to answer simple questions about object existence or count. On the other hand, reasoning models can answer questions about object relationships, comparisons, hierarchy, uniqueness etc. Recent machine reasoning models~\citep{perez2018film, mac} give an opportunity to develop new class of dialogue applications. Their higher knowledge generalization partially solves data availability problem. We consider an application for \textit{smart home} where users more freely communicate with their home appliances. We are motivated by the current home appliances~\citep{amazon} with limited number of voice commands such as turning something on or off, adjusting music volume or light conditions.

We propose to apply state-of-the-art visual reasoning model~\citep{mac} to ask a \textit{smart fridge} about its contents and various properties of the food with close-to-natural conversation experience. Our visual reasoning model is able to answer user questions about existence, count, category and freshness of each product by taking a photo using the image sensor inside the fridge. Users may chat with their fridge using a phone messenger while away from home, for example, when shopping in the supermarket. While some home appliances already can send photos of the fridge contents~\citep{samsung}, we show that our application saves user time by answering complicated reasoning questions instead of visual analysis of high-resolution photos on the small phone screen. Naturally, the next step in developing such AI applications for home appliances could be a connection of visual reasoning with more complicated tasks e.g. what to buy in the supermarket to cook a certain dish.

To demonstrate viability of our concept for smart home appliances, we implement a practical system with off-the-shelf Facebook messenger~\citep{fb} interface to communicate with the remote smart home. The messenger is connected to smart fridge through scalable cloud server. This server is responsible for answering user requests by executing computationally-challenging DNN reasoning model with text and camera snapshot modalities. We conduct initial user tests and identify typical patterns of how users tend to communicate with their smart fridge using our text interface. Based on these insights, we modify our original train dataset distribution to improve precision of the visual reasoning model.

\section{Related Work}\label{sec:related}

\subsection{Multimodal Sensing for Smart Homes}
Any model can be considered multimodal if its inputs come from different modalities e.g. a camera and laser radar for robot vacuum cleaners to perform autonomous navigation~\citep{Xu2017PointFusionDS}. It is nontrivial to build such models using feature engineering paradigm. In case of ML with learnable features, sensor fusion can be accomplished by a simple feature concatenation~\citep{Chen2016Multiview3O}. Other methods consider using modality-specific preprocessing layers or explicit regularization terms in the loss function~\citep{radu}.

A plenty of sensors can be installed in a smart home: utility meters, door access sensors, security or pet monitoring cameras etc. They can be used to accomplish a variety of useful tasks: energy saving, creating comfortable conditions, activity detection and many others. Conventional ML models are trained to perform these tasks without considering a user interface. On the other hand, VQA model can be viewed not only as a multimodal model with text input being a separate modality but also a model with explicit user interface. Then, it can be combined with a plurality of home sensors in a single end-to-end trainable ML model and provide natural-dialogue interface to interact with these sensors.

\subsection{Visual Question Answering}
VQA field is progressing rapidly with the availability of public datasets and advances in DNN models. For example, accuracy for the popular VQA dataset~\citep{vqa2} grew from approximately 55\% in 2016 to 70\% in 2017 and currently surpassed 75\% getting closer to the human-level accuracy of 81\%.

This progress in accuracy relies on advances both in vision feature extraction and language understanding. Improvements in vision include deeper residual networks~\citep{he}, use of object detection features~\citep{frcnn}, and, recently, research is heading into graph-based networks to learn explicit relational representations~\citep{Hu2019LanguageConditionedGN}. We employ deep residual DNN (ResNet101) for visual feature extraction. Such features, when pooled from the last layers of the classification network, contain information about object properties and spatial position at the scene.

The success in language understanding can be attributed to recurrent networks such as long short-term memory (LSTM) by~\citet{Hochreiter} and its evolution e.g. bidirectional LSTM (biLSTM). The recurrent networks use as inputs an embedding vectors~\citep{word2vec,glove} encoded from the sentence words. Recent BERT~\citep{bert} model is able to process the whole sentence or even paragraphs of text to extract semantic concepts at the expense of higher data and computation processing. In addition, state-of-the-art models use explicit self-attention mechanisms~\citep{allattention, Yu_2019_CVPR} to propagate only the relevant information. In this paper, we use the model that incorporates recent advances in attention mechanisms and biLSTM language encoding.

\subsection{Interfaces for Smart Home Appliances}
In past few years, home appliances are evolving from offline machinery controlled by a remote to cloud-connected devices with software application control. The concept of an interface for such software applications is still predominantly based on windows with menus and buttons.

With the success in language understanding, the concept of intelligent interfaces quickly shifting towards text or voice-controlled devices. For example, Google Assistant~\citep{google} and Alexa~\citep{amazon}, embedded into home smart hubs, offer voice input options. Such smart hubs can be connected to any home smart appliances. As of now, most of them are limited to simple on/off or up/down commands and are not able to execute reasoning tasks. 

We implement new conversational interface for smart fridge as a proof-of-concept application. The advantage of this use case is a controllable and mostly static vision environment. Some other existing products~\citep{samsung} use notion of smart fridge, but their interaction part is limited only by a feature to send photos of the fridge contents. In contrast, we propose a dialogue system where users can ask more sophisticated questions about contents of the fridge.

\begin{figure*}[t]
	\centering
	\includegraphics[width=0.9\textwidth]{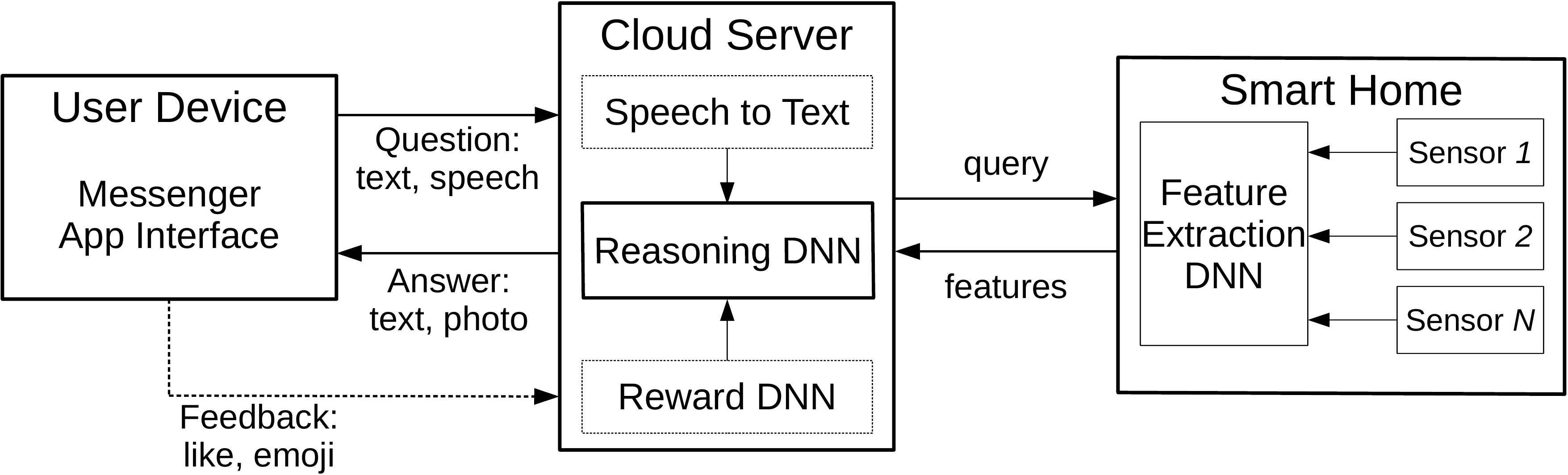}
	\caption{General concept of the proposed conversational interface with smart home appliances.}
	\label{fig:0}
\end{figure*}

\section{General Concept}\label{sec:concept}
A general concept of the proposed conversational interface with smart home appliances is depicted in Figure~\ref{fig:0}. The concept consist of three main parts: a personal device with user interface, cloud server that accomplishes processing routines, and a smart home equipped by smart appliances with multimodal sensors.

\subsection{Smart Home}
An idea of smart home has been circulated for many years, but only recently it started to catch traction in real products. Usually these products are either directly connected to Internet or through a special smart hub. Each of them has been developed separately with a predefined software command interface between application and a set of sensors. This constrains these products from execution of a more complicated abstract commands and prevents them from aggregation of multimodal sensors.

On the other hand, ML systems can offer a solution by training \textit{end-to-end} reasoning model with multiple sensor inputs. In this case, learned feature extractor converts sensor data into a model-specific representation. End-to-end learning not only improves model accuracy, but also anonymizes private data and compresses it for efficient network transfer. The drawback of this approach is lack of interpretability~\citep{explain}. Fortunately, research on explainability~\citep{ancona2018towards} and feature disentangling~\citep{achille} may lead to a more transparent ML models.

\subsection{Cloud Server}
A cloud server is responsible for communication between user device and smart home appliances. Cloud server can be distributed for robustness and to decrease round-trip question-answer latency. The second function is to run computationally-challenging DNN models upon requests from users. These requests can be in the form of text or speech audio. The latter requires an additional speech-to-text model to produce text or, preferably, can be combined with reasoning model by learning features from the raw audio.

The reasoning model receives both the sensor and user request features. An answer is calculated and returned to the user in the form of text, speech or photo. Photo answer may represent an object of interest e.g. zoomed photo of a child or pet can be returned from the monitoring cameras. In case of smart fridge, it could be a photo of the requested product.

The last component of our concept is a \textit{reward} DNN. It is responsible for acquiring user feedback in the form of likes and emojis, which are widely accessible in most of the modern messengers. This feedback can guide reasoning model to produce more relevant answers by learning user preferences. Currently, this is the most challenging part because user feedback is usually sparse and biased. The reward DNN can potentially be built using reinforcement learning~\citep{sutton}, however, reinforcement learning is known to have low sample efficiency. Realistically, only large amount of user feedback may lead to success in development of this element.

\subsection{User Device}
Users may use their personal devices e.g. phones, tablets, etc. to chat with their home appliances using a software application. Development of multi-platform application with custom interface can be long and expensive. Hence, we propose to use existing messenger applications~\citep{fb,telega}. Only text interface and an option to attach recorded speech using a microphone or to paste camera photo are needed to provide natural conversational experience. User feedback for the reward DNN can be collected using emoji and like responses. Modern messengers support such interface features and provide application programming interface (API) to connect a developed bot with a custom cloud server. Hence, users share the same interface with their friends, relatives and home appliances.

The drawback of chat bot interface is lack of buttons to perform the most frequent actions quickly like in traditional interfaces. Additionally, any unexpected model answer may lead to user frustration. The first problem can be addressed by the virtual menus, which are supported by some messengers e.g. Facebook~\citep{fb}. The model errors are the hardest to address. In our smart fridge application, we propose to send a link to sensor snapshot such that users can manually verify ambiguous answers.

\section{FRIDGR Dataset}\label{sec:dataset}
Due to lack of public datasets for home appliance applications, we generate a synthetic dataset called~\textit{FRIDGR}. FRIDGR associates fridge objects with the corresponding text questions and answers through randomization of object appearances and views. FRIDGR is produced by analogy to popular CLEVR~\citep{clevr} dataset with similar annotation format and software scripts. We make publicly available the scripts to reproduce FRIDGR or to construct another task-specific application.

\begin{figure}[t]
	\centering
	\begin{tabular}{cc}
	\includegraphics[width=0.45\columnwidth]{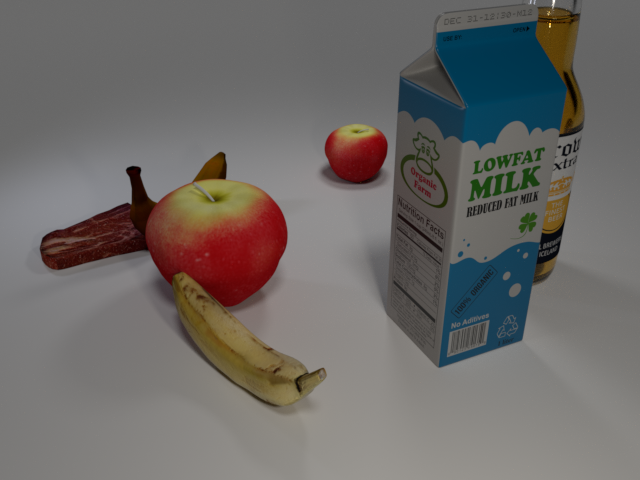} &
	\includegraphics[width=0.45\columnwidth]{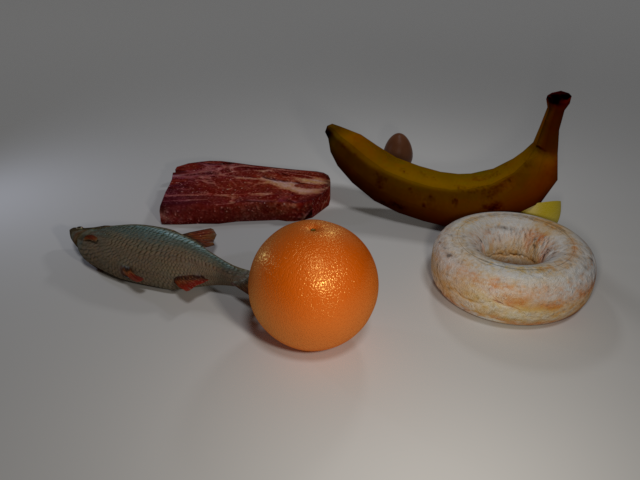}
	\end{tabular}
	\caption{Examples of synthesized FRIDGR images.}
	\label{fig:im}
\end{figure}

\subsection{Photo-Realistic Image Generation}
Unlike diagnostic CLEVR dataset, we are interested in synthesizing photo-realistic images that can generalize to real-world objects. Therefore, we use 3D models with textures from real objects to decrease gap between synthetically-generated models and real physical objects. This gap can be decreased not only by producing realistic images, but also using additional methods from domain adaptation~\citep{Wang2018DeepVD} field. With availability of unlabeled real data and synthetically generated annotated examples, domain adaptation allows to train a model that performs well on the real data.

Another potential issue is the large distribution of real objects. For instance, fridge products may look very different from one country to another. Then, train dataset has to be customized for each geographic region to achieve positive user feedback. For proof-of-concept demonstrations we propose to finetune the model trained on generated images using a small dataset of manually annotated images. Such real images have to come from the distribution of physical objects planned to be used during interactive demonstrations.

Overall, FRIDGR contains 60,000/10,000/10,000 images in training, validation and test datasets, respectively. The generation process is done using the Blender tool~\citep{blender} and graphics processing units (GPUs) to decrease processing time. It took approximately 4 days for image generation using 8 P100 GPUs. Examples of FRIDGR images are shown in Figure~\ref{fig:im}. FRIDGR dataset consists of food products with 14 classes from 5 categories and certain properties such as size and freshness.

Each image contains a set of generated objects imposed on a fridge shelf. Every object has a label with true bounding box coordinates, its category, freshness and size properties. In addition, object relationships can be encoded explicitly into ground truth format~\citep{genome}. The following types of objects and their properties are supported:
\begin{itemize}
	\item Classes: donut, coke can or bottle, beer, apple, banana, lemon, orange, pear, egg, meat, milk, tomato and fish.
	\item Categories of each class: dessert, drink, fruit, vegetable, ingredient.
	\item Freshness: fresh or expired. Freshness property can be applied to apple, banana and meat.
	\item Size: each object comes in small or large size.
\end{itemize}

\begin{figure*}[ht]
	\centering
	\includegraphics[width=0.64\textwidth]{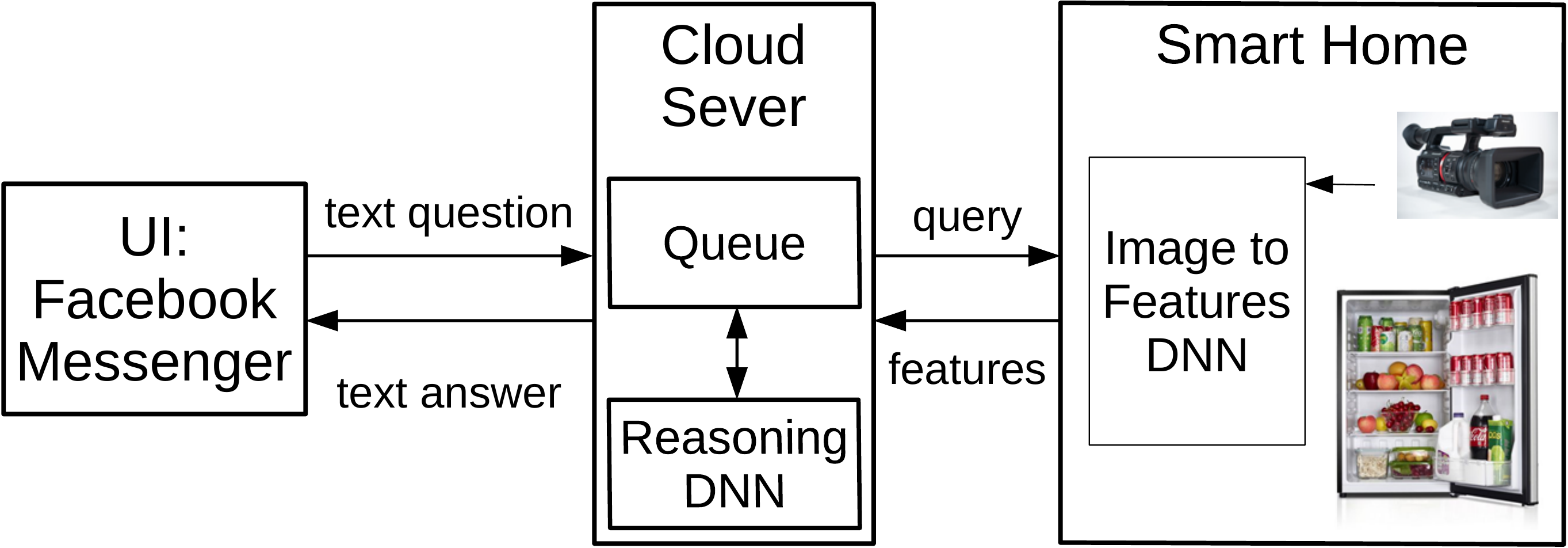}
	\caption{Diagram of the developed system.}
	\label{fig:2}
\end{figure*}

\subsection{Generation of Question-Answer Pairs}
The vision part of the dataset is constrained by the distribution of real objects and their scene appearance. On the other hand, the question answering part mostly depends on the application and user experience requirements. The expectations about a dialogue system cannot be clearly identified without actual user tests. Here we concentrate on a technical part of dialogue ground truth generation and describe user tests and insights in Section~\ref{sec:evaluation}.

We compose synthetic questions using language templates. Each template generates a single question type. Due to language variability every semantically-similar type may have various syntactical forms. We explicitly encode plurality of these forms into each template. It is important to cover all possible combinations of user questions. For example, template for so called \textit{existence} questions is defined as follows:
\begin{itemize}
	\item Are there any $\mathrm{<Z> <M> <C> <S>}$?
	\item Any $\mathrm{<Z> <M> <C> <S>}$?
	\item Is there $\mathrm{<Z> <M> <C> <S>}$?
	\item Do I have $\mathrm{<Z> <M> <C> <S>}$?
\end{itemize}
Template variables are defined as $\mathrm{<Z>}$ - size, $\mathrm{<M>}$ - freshness, $\mathrm{<C>}$ - category, and $\mathrm{<S>}$ - class. Each of these variables is not compulsory because we intend to generate a distribution of all possible hierarchical relationships. This is implemented using random masks to enable or disable certain variable. For instance, we can generate full question: "do I have large fresh banana?". Or we can exclude category and class variables and ask: "do I have fresh products?". The former targets specific objects in the hierarchy tree, while the latter reasons about broad range of items inside the fridge. We implement special routines to avoid tautology questions e.g. "do I have fresh fruit bananas?". In addition, we substitute typical subjects e.g. "products" or "items" in cases when generated mask does not have any subject. Lastly, each word is randomly replaced by its synonyms. For example, "drink" category has "beverage" and "soda" synonyms as well as its plural forms. 

Each image is accompanied by approximately 30 randomized question-answer pairs written to scene representation file along with object ground truth bounding boxes and their relationships. The questions are asked not only about present but also absent objects with negative answers. In total, FRIDGR dataset contains 1.8/0.3/0.3 million question-answer pairs in training, validation and test datasets, respectively. Currently, the following types of questions are supported:
\begin{itemize}
	\item Existence: is there an object of this class?
	\item Count: how many objects of this class?
	\item Category: users can use object categories instead of classes such as drinks, desserts, fruits, vegetables etc.
	\item Freshness: subset of objects may have darker color which signals about their expiration e.g. dark meat or banana.
	\item Size: each object comes in small or large size.
	\item Combinations of properties: any combination of the described properties e.g. "how many large fresh bananas" or "how many fruits".
\end{itemize}

\section{The Developed System}\label{sec:system}
The reference concept of the proposed system has been introduced in Section~\ref{sec:concept} and shown in Figure~\ref{fig:0}. In this section, we describe the key elements of the implemented system. The cloud server side code is written in Node.js~\citep{nodejs} and ML part is implemented in Python using popular TensorFlow~\citep{tf} and PyTorch~\citep{paszke2017automatic} DNN frameworks.

\subsection{User Interface}\label{sec:ux}
\begin{figure}[t]
	\centering
	\includegraphics[width=0.28\columnwidth]{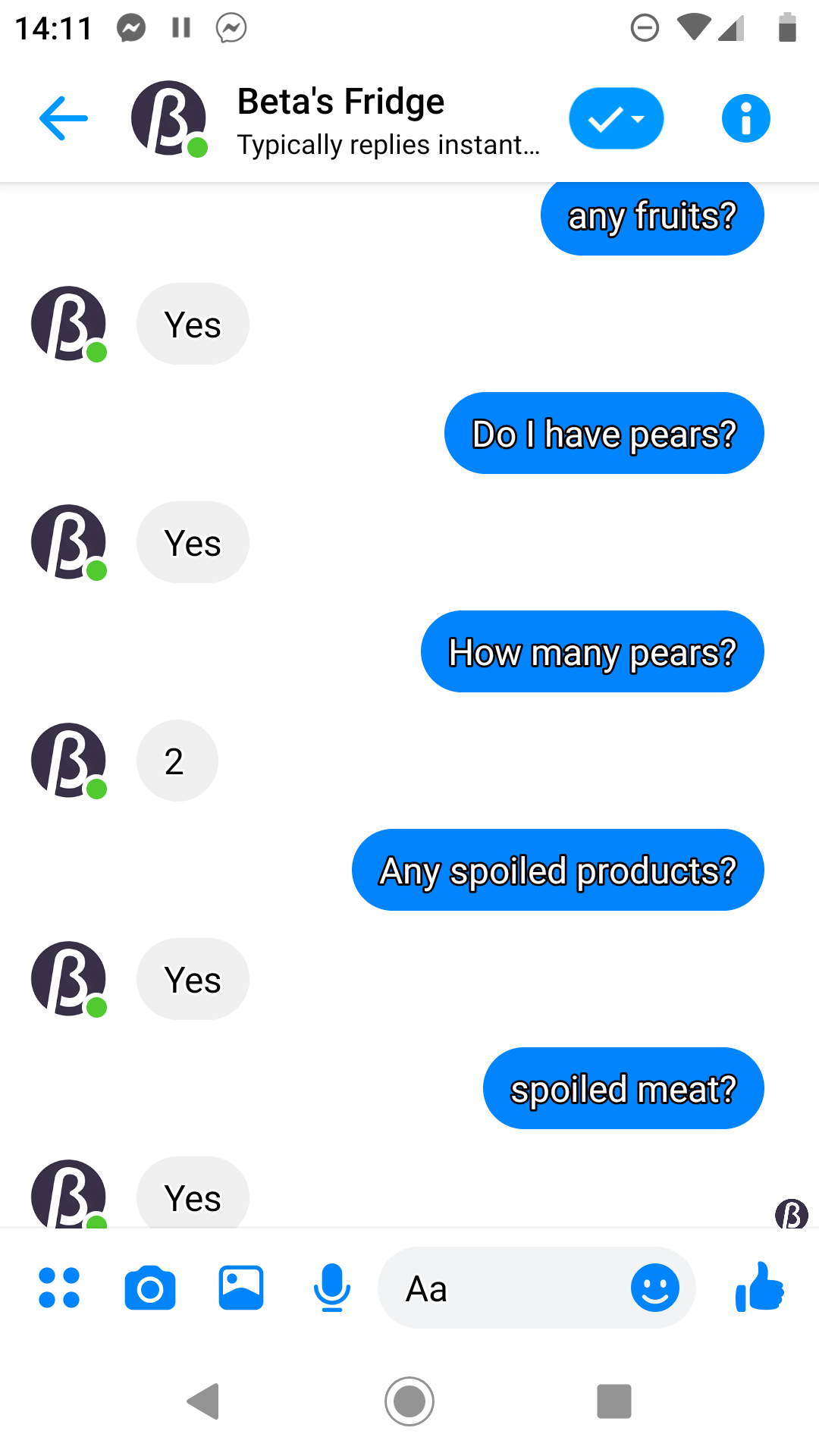}
	\caption{Messenger's user interface and example of the dialogue between user and fridge.}
	\label{fig:3}
\end{figure}

A number of messengers provide custom chat bots. Currently, we support only Facebook messenger, though any other one can be connected to cloud server. The interface of this messenger has all the required features: text and microphone entry, camera attachments, like and emoji responses. In addition, it has features to draw menus pushed from the cloud side and to embed thumbnail photos. An example of the realized interface is shown in Figure~\ref{fig:3}. In future, we plan to accompany text answers with the fridge snapshots zoomed at the object of interest. 

\subsection{Cloud Server}\label{sec:cloud}
Our cloud server runs on Heroku server~\citep{heroku} and contains two main parts. First, all user requests are pushed into a serving queue and indexed by a unique identification number. Then, the server sends a command to smart home fridge camera to capture current photo. This photo is processed through DNN feature extractor and feature vector is sent back to cloud server. Once the sensor data arrives, cloud server runs our reasoning DNN model for queued requests. Our infrastructure with the serving queue allows to serve multiple users at the same time. DNN models are typically executed on GPUs, which are able to efficiently process a \textit{batch} of requests at a time from the queue.

Compared to the proposed concept, we did not implement end-to-end speech-to-text model. Speech recognition may suffer the problems with noisy environment and lack of adaptation to a particular user voice. However, speech recognition can be quickly added using existing cloud services e.g. Google API~\citep{googleapi}. We leave the proposed reward DNN concept as future research direction due to theoretical and practical difficulties described in Section~\ref{sec:concept}.

\subsection{Visual Reasoning Model}\label{sec:model}
\begin{figure*}[t]
	\centering
	\includegraphics[width=0.94\textwidth]{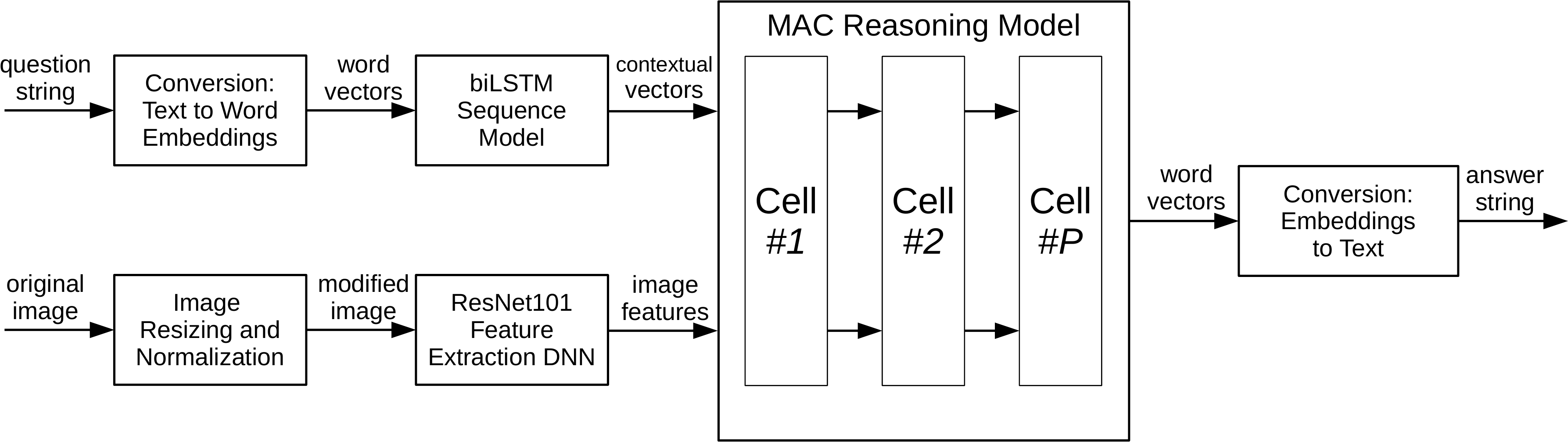}
	\caption{Diagram of the implemented MAC reasoning model.}
	\label{fig:mac}
\end{figure*}

A schematic diagram of the visual reasoning model is presented in Figure~\ref{fig:mac}. An input text question is divided into a sequence of words. Each word is converted into embedding vector of size $300\times1$. Produced embeddings are initialized with widely-used pretrained GloVe~\citep{glove} vectors. Next, question vectors are passed to bidirectional LSTM to learn sequential relationships, which generates contextual vectors of size $512\times1$. Bidirectional modeling allows us to find order-invariant dependencies. 

Similar process of input conversion into feature vectors happens for camera images. Images are preprocessed into fixed $224\times224$ resolution and normalized to zero-mean unit-variance format. Then, the pretrained ResNet101 extracts representation of size $1024\times14\times14$, where $14\times14$ denotes spatial resolution and $1024$ is number of per-location features. The dimensionality choice depends on the total number of classes and maximum amount of objects to analyze. It might be important to increase these dimensions for practical fridge application and to concatenate images from multiple views for occluded objects.

Finally, both feature modalities are passed to MAC architecture, which contains $P$ identical cells as depicted in Figure~\ref{fig:mac}. Each cell extracts object concepts iteratively, passes this information from cell to cell, and the final cell produces an answer. The low-level concepts are learned by the first cells and a more abstract reasoning happens in subsequent cells in a compositional manner. We do not provide the very details about the MAC architecture which can be found in~\citep{mac}. At the same time, we experimented with several architectural choices to achieve the best results. For example, we noticed that explicit self-attention and memory gating helps to improve accuracy results. Lastly, we point out that MAC architecture allows to visualize iterative reasoning process using both linguistic and visual attentions masks. This significantly improves model interpretability.

\section{Experiments}\label{sec:evaluation}
\subsection{Demo Setup and Interaction Setting}\label{sec:demo}
The implemented messenger application has been installed on iPad tablet and presented to users. Before starting user tests, we described FRIDGR dataset objects, their properties and what kind of questions can be potentially asked. To familiarize users with the basics of our scene representation, we showed visual examples of the fridge photos. We conducted user tests with two separate groups of participants. The first group has been verbally instructed about the details of user interaction with the demo setup. The second group has seen only the poster with common question patterns and FRIDGR images.

One of the goals is to check whether the response time is acceptable to users. A typical recorded response time is within a second range for our setup. The reasoning model along with all data preprocessing on the server side contributes only approximately 100 milliseconds of latency with GPU processing. Approximately 80\% of the latency (few hundreds of milliseconds) is introduced by communication between user device and two consecutive servers. An additional penalty is introduced by the trip between the Facebook chat bot server and our cloud server. It can be avoided if combine both functions into a single cluster. We conclude that users did not experience any difficulties related to interface latency.

\subsection{User Tests Assessment and Insights}\label{sec:hci}
We conducted user tests with approximately 30 participants in both groups. The first group consists of 5 persons, while others belong to the second \textit{less trained} group. We showed 3-5 random visual examples to the first group and 1-2 examples to the second group. Each participant asked from 5 to 20 questions. In total, we collected approximately 450 questions.

We identify several patterns of user interaction with our interface. First, participants tend to ask questions in a very short form to decrease typing time. For example, in place of longer question "do I have any apples", they prefer a short forms e.g. "any apples?". This is very common pattern that can be found in 65\% of questions. We did not take into account this pattern in our original FRIDGR dataset. Second, sometimes participants ask incomplete one-word queries e.g. "milk?". Then, it is not clear what was meant by this question. Most of the time, they presume the existence question, which should be reflected in the question templates. Third, users usually avoid using question marks and articles. Issues with this pattern can be avoided during preprocessing step.

In general, we do not notice any significant difference between more trained first and less trained second groups. We incorporate the insights from user tests into our \textit{modified} FRIDGR dataset. The vision part of the modified FRIDGR is identical to the original dataset. We change only question templates and extend them by adding identified short forms of questions. Also, we modify distribution of variable masks from Section~\ref{sec:dataset}. Instead of generating long questions about specific objects e.g. "are there any small fresh bananas there?", we produce mostly short questions that presume certain hierarchical reasoning e.g. "any bananas?" or "any fruits?". This helps to emphasize common user questions during training of our ML model. Lastly, we modify preprocessing step to remove semantically unimportant symbols such as articles, question marks, and commas.

\subsection{Quantitative Model Accuracy}\label{sec:quant}
We train the original MAC~\citep{mac} models adopted for FRIDGR data. We employ two reference models: model\#1 with $P=4$ cells and model\#2 with $P=6$ cells. They differ not only in the number of cells, but also in architectural choices. Model\#2 adds control-based gating mechanism over the reasoning memory. As part of ablation study, we train these models to identify accuracy bottleneck, which can be related to either model complexity or dataset distribution.

\begin{table}[htbp]
	\caption{Accuracy of the original and modified FRIDGR datasets.}
	\begin{center}
		\begin{tabular}{ccccc}
			\toprule
			{} & \multicolumn{2}{c}{Original} & \multicolumn{2}{c}{Modified}\\
			{} & Train,~\% & Test,~\% & Train,~\% & Test,~\%\\
			\midrule
			Model\#1 & 99.19 & 93.67 & - & 60.07\\
			Model\#2 & 99.95 & 94.95 & - & 55.55\\
			\midrule
			Model\#1 & - & - & 98.59 & 95.18\\
			Model\#2 & - & - & \textbf{98.70} & \textbf{95.38}\\
			\bottomrule
		\end{tabular}
		\label{tab:1}
	\end{center}
\end{table}

We present FRIDGR accuracy on train and test datasets for both models in Table~\ref{tab:1}. The third column results show that the larger model\#2 achieves approximately 1.2\% higher accuracy on test data compared to the model\#1 for the originally generated dataset: $94.95\%$ and $93.67\%$, respectively. Therefore, we conclude that the model complexity is not a bottleneck for this dataset and almost all test questions can be successfully answered.

Next, we report accuracy results for the user-guided modified FRIDGR dataset. Note that the vision part as well as question templates with synonym vocabulary are identical for both the original and modified datasets. The last column of Table~\ref{tab:1} shows that even relatively small distribution shift in language part may lead to significant ($35-40\%$) drop in test accuracy. This result highlights the importance of data selection, which has to cover all possible semantic templates and variability in language constructs. In our case, we achieve this using human-in-the-loop guidance with subsequent question randomization.

\begin{figure*}[t]
	\centering
	\includegraphics[width=0.98\textwidth]{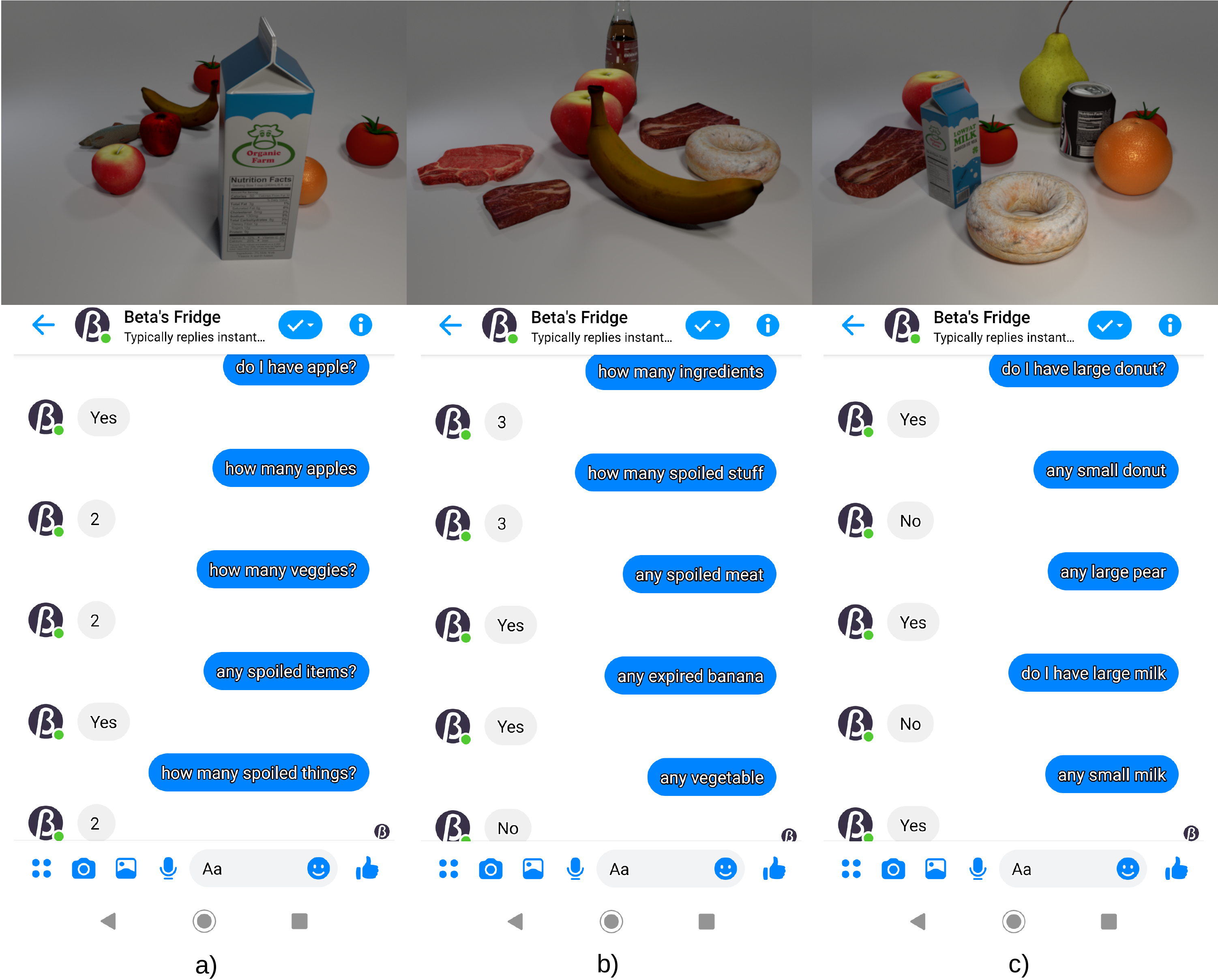}
	\caption{Examples of dialogue using the proposed system.}
	\label{fig:qual}
\end{figure*}

\subsection{Qualitative Results}\label{sec:qual}
In this section, we present qualitative results to demonstrate system abilities. We picked three illustrative visual examples from the test dataset. Important to note that we did not finetune model for these particular examples. We asked a typical questions about each of them in the same user interface as during tests. These dialogue examples are shown in Figure~\ref{fig:qual}.

Figure~\ref{fig:qual}(a) contains multiple fridge objects and some of them are partially occluded (tomato and orange). The are two expired items: dark banana and dark red apple. Then, we ask questions about count or existence of particular objects as well as their categories and freshness. Notably, the visual reasoning model is able to distinguish between all these abstract properties e.g. "how many veggies?" and "how many spoiled things?". Compared to explicit vision detection systems, this question-driven reasoning effectively encodes numerous possible relationships, which are growing exponentially with the number object properties.

Figure~\ref{fig:qual}(b) scene has three pieces of meat (ingredient category), where two of them are expired. Also, there is one spoiled banana (fruit category). Then, the system correctly answers the question about "how many spoiled stuff", even though objects belong to different categories. In addition, it correctly answers "any vegetable" question with no objects of this category.

Figure~\ref{fig:qual}(c) scene mostly represents the questions about size of the objects. Our model correctly answers questions about large donut in front, small milk pack and giant partially occluded pear in the back. The questions of this type rely on accurate vision sub-system, where the abstract size property is implicitly learned from the data labels. 

\subsection{Interpretability}\label{sec:interp}
\begin{figure}[t]
	\centering
	\includegraphics[width=0.64\columnwidth]{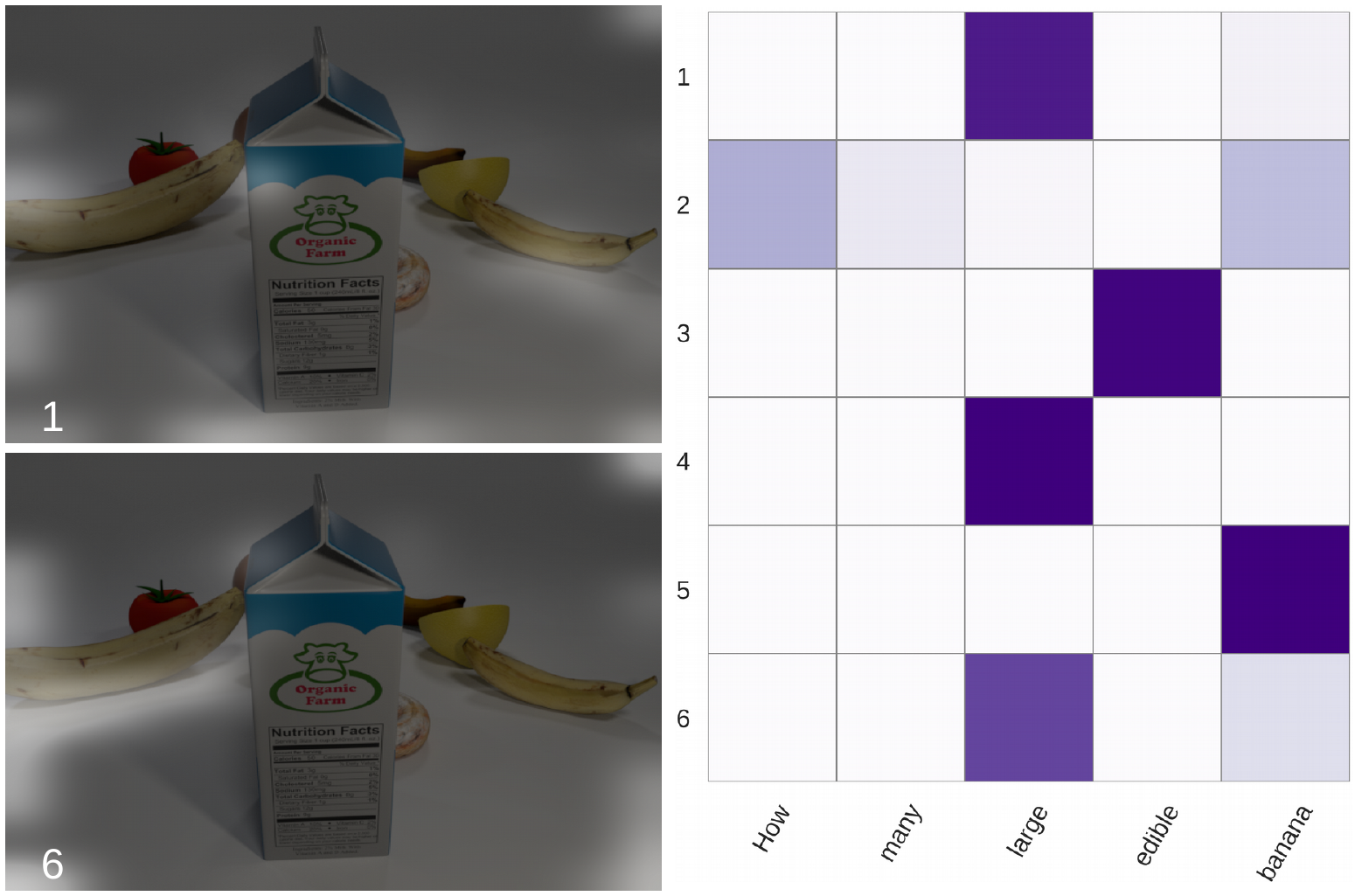}
	\caption{Example of visual and question attention masks: visual attention mask for cells $P=1,6$ and question attention for all $P=1\ldots6$ cells.}
	\label{fig:interp}
\end{figure}

An important feature of MAC visual reasoning model is its compositional iterative architecture with $P$ identical cells. Then, the attention masks produced by each cell can help to understand how concepts are learned after each iteration. We use our trained model\#2 with $P=6$ and pick an example shown in Figure~\ref{fig:interp}.

The question attention masks illustrate what concepts are mostly learned at each cell. First, our model attends to "large" and "banana" features and "how many" question type. Then, a more abstract "edible" property is taken into account, and, finally, the model focuses again on "large banana". This may explain why the smaller model\#1 with $P=4$ cells performs almost the same as model\#2: the short questions can be processed by a less number of reasoning steps.

In addition, we show visual attention masks for the first (top) and last (bottom) cell in Figure~\ref{fig:interp}. The first cell is unable to attend to the correct answer yet, which is a large banana on the left corner. Instead, it focuses on all bananas and an intersection of multiple objects. The last cell attends to the correct object on the left and rejects the small banana on the right side as well as the dark (inedible) occluded banana in the background. Therefore, this attention mask visually explains the correct answer.

\section{Conclusion and Future Work}\label{sec:conclusion}
In this paper, we introduced the concept of conversational interface between users and smart home appliances. We described its main components including the crucial multimodal machine reasoning model. Smart fridge has been selected as a proof-of-concept application. The proposed visual reasoning model realized a dialogue system where the users are able to ask questions about visual contents of the smart fridge. This improves user experience by saving their time when using personal devices with small screen or when these devices cannot be accessed. The key feature of our concept is its ability to answer more complicated reasoning questions compared to current commercial interfaces.

We reviewed difficulties and potential solutions during engineering of such ML models including data availability and model selection aspects. To overcome the data availability issue, we developed application-specific synthetic dataset called FRIDGR. We described the ways to accomplish knowledge transfer to real physical objects using either domain adaptation methods or finetuning with the small annotated dataset. The selected MAC model achieved more than 95\% accuracy on the challenging FRIDGR test dataset with high complexity and variability of questions types. In addition, our model was able to process such questions in iterative compositional way with interpretable attention masks during reasoning process.

To demonstrate our concept in practice, we implemented the key elements of the proposed system. We employed the existing messenger and showed that its off-the-shelf interface is suitable for communication with smart home appliances using natural dialogue unlike traditional menu-based applications. The messenger was connected to the developed cloud server back-end, which is able to serve multiple users in parallel and execute computationally challenging DNN routines in a distributed way.

We conducted initial user tests using our demonstration setup. Experiments showed that people prefer to ask very short or even incomplete questions to increase interaction speed. Quantitative experiments showed a significant drop in accuracy for ML models when the distribution of user questions diverges from the training dataset. Based on these insights, we modified our question generation templates and distribution of the question types in training dataset. This human-in-the-loop guidance restored 95\% test dataset accuracy.

In future, we would continue to extend the existing system by adding new types of sensors, better interface features such as sensor previews, speech and photo queries. As a future research, we envision two potential directions. First, the reward model can be added to the system to learn user preferences and to dynamically adjust reasoning model using the feedback channel. This feedback might be emulated during dataset generation step. Second direction is a new application that combines fridge contents reasoning and recipe understanding. This recipe-ingredient reasoning can be realized using recently published datasets e.g. Recipe 1M+~\citep{marin2019learning}. Imagine an application that proposes to buy certain products in a grocery store in response to the user's desired recipe and fridge contents.

\bibliography{paper_arXiv}
\bibliographystyle{iclr2020_conference}

\end{document}